\documentstyle[11pt,newpasp,twoside,epsf]{article}
\makeatletter

\def\@journalname{ASP Conference Series}
\def\cpr@holder{Astronomical Society of the Pacific}
\def\@jourvol{000}
\def\cpr@year{2002}
\def\vol@title{Hubble's Science Legacy: Future Optical-Ultraviolet Astronomy from Space}
\def\vol@author{K.R. Sembach, J.C. Blades, G.D. Illingworth, R.C. Kennicutt (eds.)}
\def\@slug{{\tabcolsep\z@\begin{tabular}[t]{l}\vol@title\\
\@journalname, Vol.\ \@jourvol, \cpr@year\\
\vol@author
\end{tabular}}
}
\def\ps@paspcstitle{\let\@mkboth\@gobbletwo
\def\@oddhead{\null{\footnotesize\it\@slug}\hfil}
\def\@oddfoot{\rm\hfil\thepage\hfil}
\let\@evenhead\@oddhead\let\@evenfoot\@oddfoot
}
\def\ps@myheadings{\let\@mkboth\@gobbletwo
\def\@oddhead{\hbox{}\hfil\sl\rightmark\hskip 1in\rm\thepage}%
\def\@oddfoot{}%
\def\@evenhead{\rm\thepage\hskip 1in\sl\leftmark\hfil\hbox{}}%
\def\@evenfoot{}\def\sectionmark##1{}\def\subsectionmark##1{}}
\def\@leftmark#1#2{\sec@upcase{#1}}
\def\@rightmark#1#2{\sec@upcase{#2}}

\makeatother

\def\edcomment#1{\iffalse\marginpar{\raggedright\sl#1\/}\else\relax\fi}
\reversemarginpar

\begin{document}

\title{Measuring the Influence of Supernovae at High Redshift}
 \author{Kurt Adelberger}
\affil{Harvard-Smithsonian Center for Astrophysics, 60 Garden St., Cambridge, MA 02138}

\begin{abstract}
Supernovae play a large but poorly understood role in our attempts
to explain the evolution of the baryonic universe.  Numerous
observations throughout astronomy cannot be explained if we
neglect their influence, yet our quantitative understanding
of the ways in which supernovae affect the universe remains
remarkably poor.  This is one of the most embarrassing
gaps in our knowledge of the cosmos, and
planned telescopes and surveys will 
probably not do much to fill it.
The problem is that these surveys will be optimized to
observe galaxies and intergalactic material independently
of each other, while (in the author's view) by far the best information
will come from simultaneous surveys of galaxies and
the intergalactic material (IGM) in their vicinity.  
Only this will show directly
how galaxies affect their surroundings and provide a rough
energy scale for supernova-driven winds.
Redshifts $1\la z\la 3$ are ideal for 
the joint galaxy/IGM surveys we advocate,
because the comoving density of star formation is near its peak,
because the
Lyman-$\alpha$ forest is thin enough for QSO spectra to reveal
the locations of the dominant metallic species,
and because bright background QSOs are common.  But a new
UV-capable spectrograph in space will be required.
\end{abstract}

\section{Introduction}
Fifteen years from now we will be awash in galaxies.  2dF, Sloan,
and maybe a Sloan successor will have given us redshifts
for more than a million galaxies in the nearby universe.
DEEP and VVDS will have added $\sim 10^5$ galaxies out to
redshift $z\sim 5$.  NGST and 30m optical/IR telescopes on the ground
may have detected thousands of galaxies to $z\ga 10$---some of the
first sources of light in the universe.  These surveys and others will
teach us a tremendous amount about galaxy and structure formation.
Our communal efforts so far will seem little more
than a prologue to the vast literature on galaxies
that will exist in 15 years.  

Nevertheless I suspect
that one of the most fundamental questions in galaxy formation
(and in all astronomy!) will remain
largely unanswered.  This is the role that supernovae played
in shaping the baryonic universe.  The influence of supernovae is
thought to account for a wide range of observations throughout
our field.  The disruption of star-formation by 
supernova explosions is the favored explanation for why so
few baryons are found in stars today (e.g., White \& Rees 1978,
Springel \& Hernquist 2002).  
Numerical simulations
cannot reproduce the large disk galaxies that we observe around us
unless they include substantial heat input from supernovae
(e.g., Weil, Eke, \& Efstathiou 1998).
The material between galaxies at high-redshift is hotter than
would be expected if gravity and the background radiation
field were the only sources of heating; another source,
presumably supernovae, appears to be required (Cen \& Bryan 2001).
It is difficult to explain why the soft X-ray background
is so faint and so dominated by AGN without asserting that
supernovae blew apart dense clumps of baryons that would otherwise
have produced copious free-free emission (e.g., Pen 1999).   
The shape of galaxy clusters' X-ray temperature/luminosity
relationship differs from naive expectations in a way that
suggests that supernovae may have imparted $\sim 1$ keV of energy to each
of the young universe's nucleons (e.g., Kaiser 1991; 
Ponman, Cannon, \& Navarro 1999).

These examples are only a few of many.  We are unable to account
for much of what we observe around us without invoking the indistinct
notion of strong
supernova ``feedback,''  and our understanding of the evolving universe
will remain seriously incomplete until we 
comprehend quantitatively how this feedback works.

The standard picture is that
the numerous supernova explosions in a young
galaxy create an enormous blast-wave (or ``wind'') that rips through the galaxy
and lays waste to its surroundings. 
But working through the details of this picture remains challenging
even after 30 years of theoretical studies.
There are many complications, but the central problem 
is that we have little idea of the characteristic energy scale
to associate with the blast-waves that supernovae drive.
The energy released by a single supernova, $\sim 10^{51}$ erg,
is known, but it is unclear how large a fraction of the energy released
by supernova explosions is imparted to nascent winds.
Much of it may be harmlessly radiated
away by the dense gas it heats.
Physical arguments and numerical simulations are at present
incapable of estimating a priori the energy of a galaxy's
wind to within even an order of magnitude---and the energy of
the winds is largely what determines how large an impact they
have on the evolving baryonic universe.
For this reason it is still unclear (e.g.)
which sorts of galaxies were responsible for seeding the
intergalactic medium with metals, or what effect blast-waves
have on galaxy formation and evolution, or even whether
realistic blast-waves would be physically capable of filling
the large role that they are assigned in the standard lore.

Whether numerical calculations in 2017 will be able to reliably
estimate the energy of galaxies' winds from first principles
is anyone's guess, but in any case we will certainly
want empirical support for the numerical results.
Properties of galaxies (e.g., disk sizes) and of the IGM (e.g., metal content)
are affected by the strength of supernova winds, and
I can imagine convoluted and uncertain chains of reasoning
that would provide an estimate the strength of winds
from observations of one or the other;
but surely
the most straightforward way to estimate the strength of 
galaxies' winds is to observe galaxies and the intergalactic
medium simultaneously and see how much winds have
disturbed galaxies' surroundings.  Winds
lose energy as they climb out of galaxies' potential wells
and crash into nearby intergalactic matter, and a robust
(if crude) measurement of the typical energy of galaxies' winds can be obtained
by seeing how far they are able to propagate.  
Simultaneous observations of galaxies and the intergalactic
material that surrounds them are easy to obtain,
at least in principle:  one need only conduct a galaxy redshift survey in a field
that contains background QSOs whose absorption spectra reveal
the locations of HI and metals in the IGM.  These sorts of
observations, present and future, are the subject of my talk.


%

\section{Current data}
\begin{figure}
\plottwo{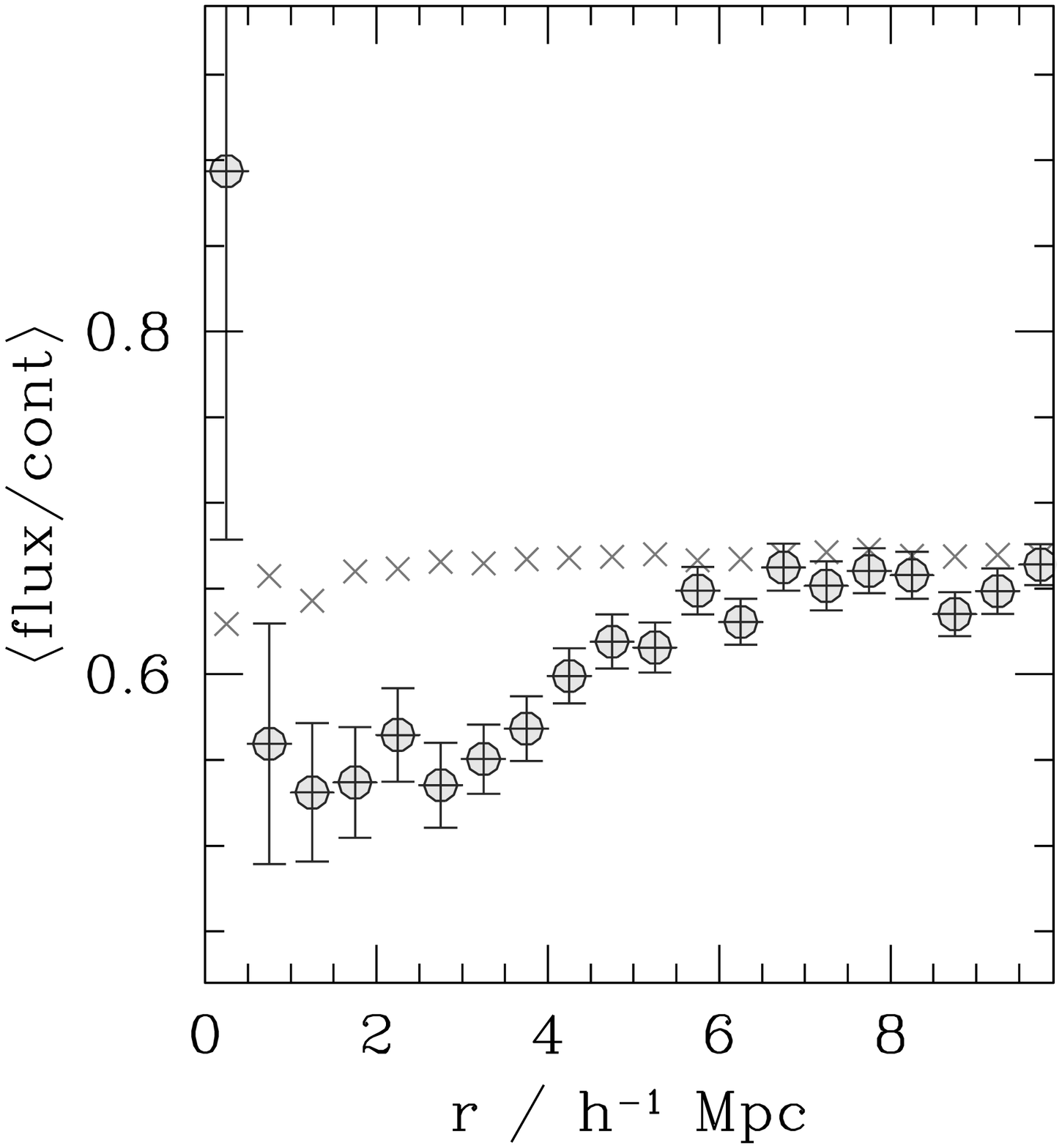}{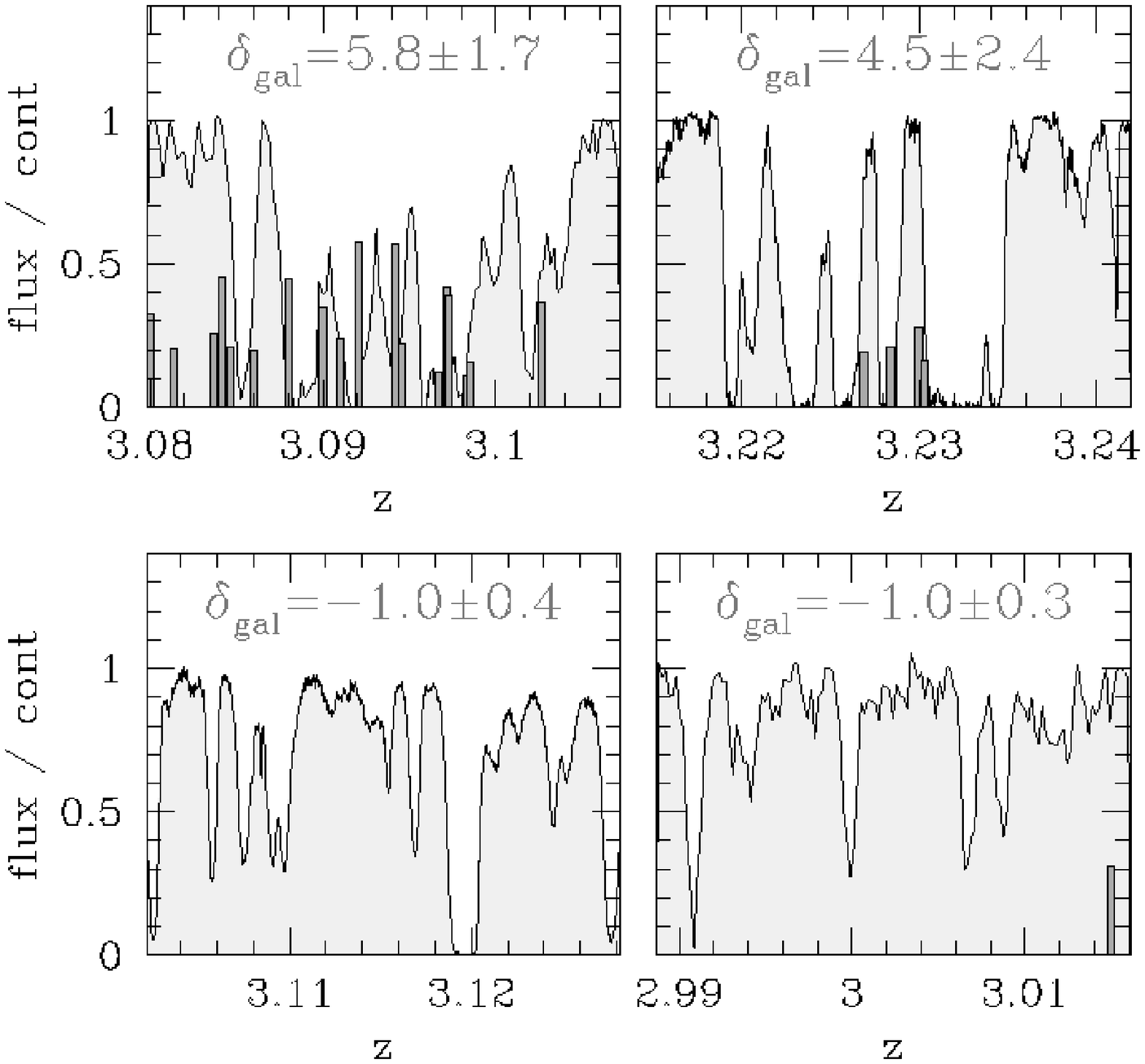}
\caption{Left:  the mean Lyman-$\alpha$ transmissivity of the IGM as a function
of comoving distance from Lyman-break galaxies.  Points with error bars
show our measurements; crosses show the result if we randomize the galaxy
redshifts.  Since Ly-$\alpha$ transmissivity decreases as the HI content of
the IGM is increased, the plot shows that Lyman-break galaxies tend to
be surrounded by large amounts of HI at comoving radii $1\la r\la 5h^{-1}$ Mpc
but little HI at smaller radii.  Right:  The HI content of the IGM along skewers
through our two most significant galaxy overdensities (top) and underdensities (bottom).
The shaded region shows the Ly-$\alpha$ absorption in the QSO spectrum; vertical
bars mark the measured redshifts of galaxies, which are typically a few comoving
Mpc away from the QSO's line-of-sight.
The overdensities, presumably young Abell clusters, contain large amounts
of HI---a result consistent with the idea that gas between the galaxies has
been roiled by supernova-driven winds.  See Adelberger et al. (2002), Adelberger (2003).
}
\end{figure}

\begin{figure}
\plottwo{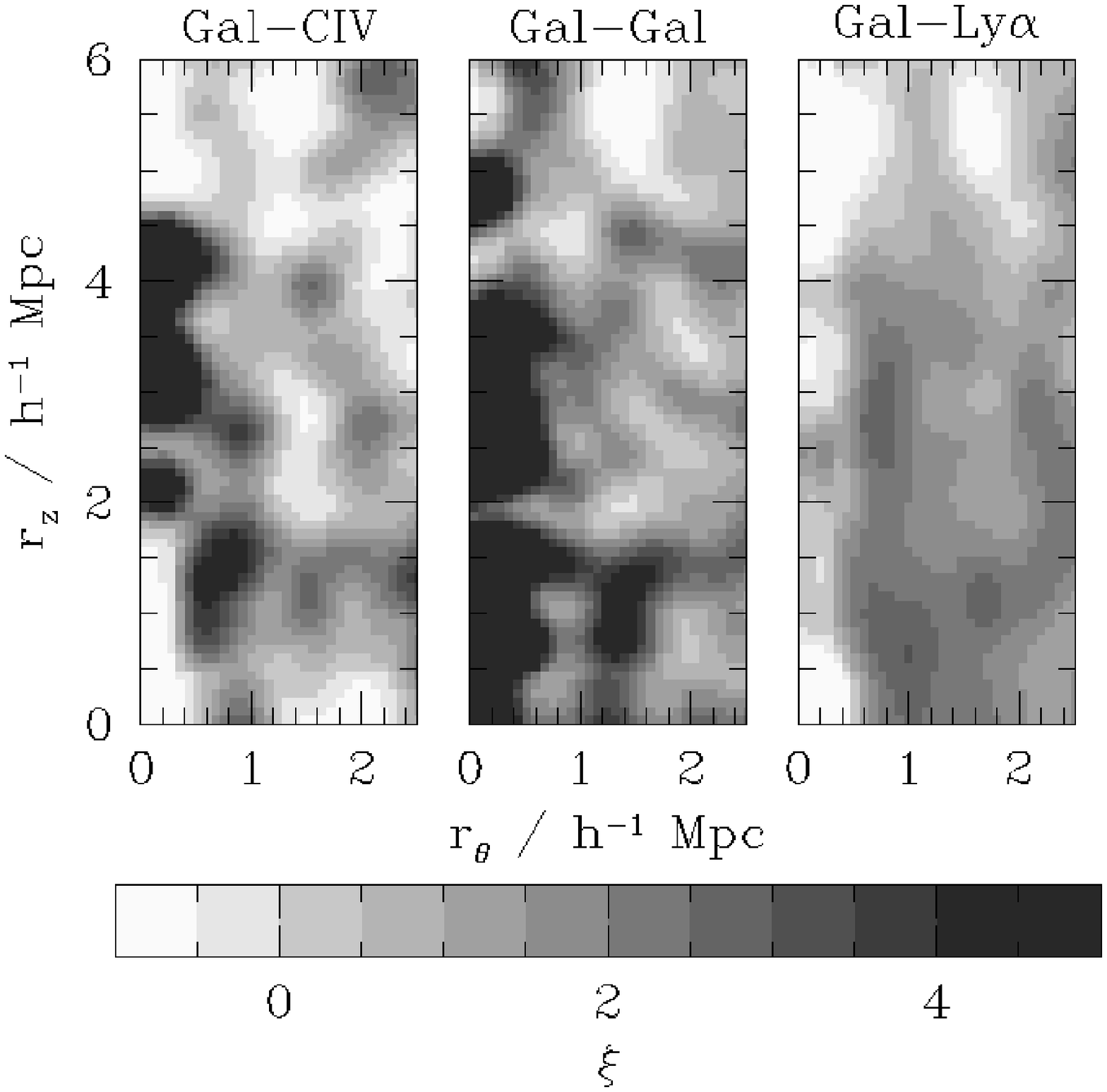}{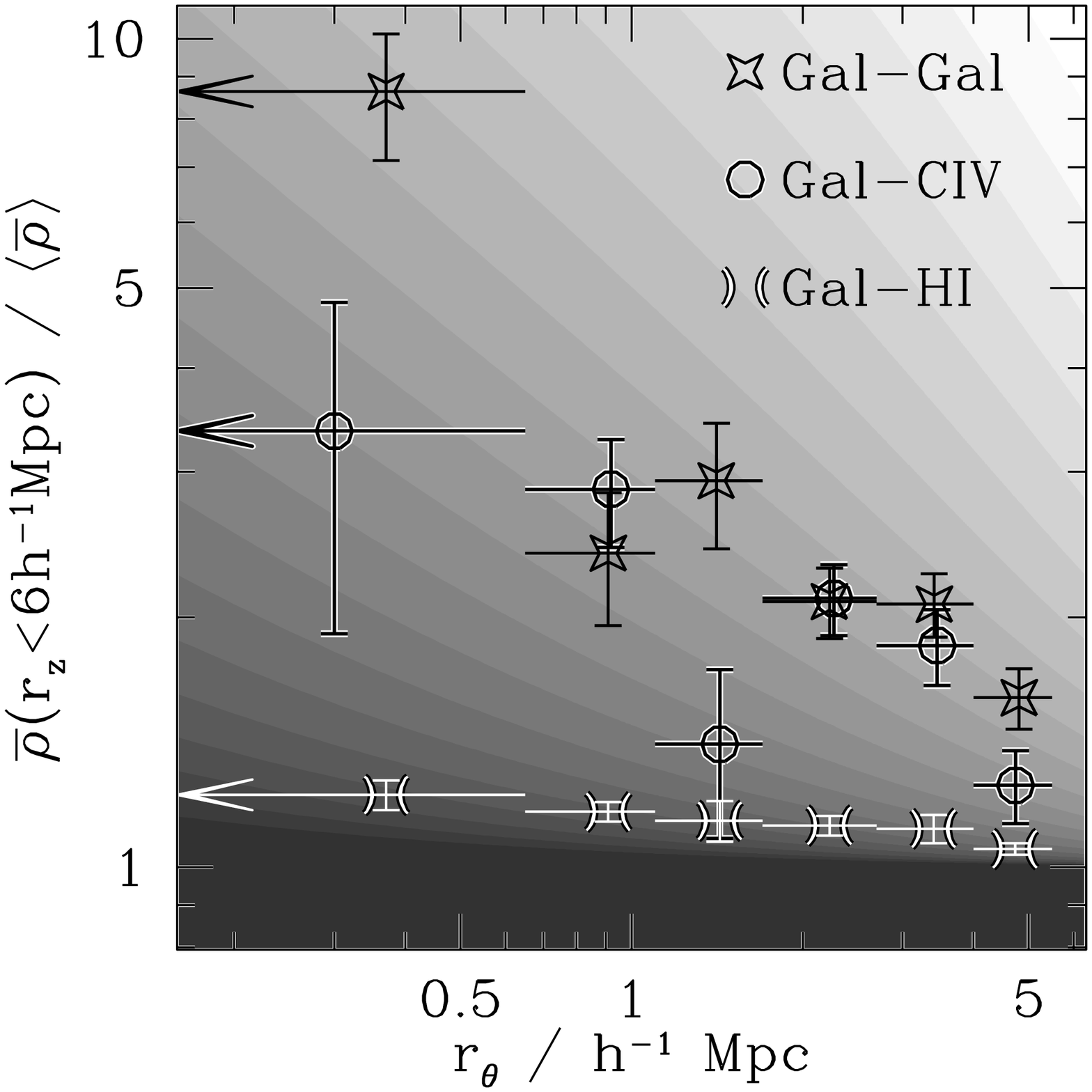}
\caption{Left:  the two-dimensional correlation functions of galaxies with
galaxies, with CIV absorption systems, and with Lyman-$\alpha$ transmissivity
decrements.  
Distances along the line-of-sight ($r_z$) and along the
plane of the sky ($r_\theta$) are comoving.  The Ly-$\alpha$ panel was multiplied
by $-10$ to make its structure visible.  
Right:  a marginalization over redshift of the data in the left panel.
Points with error bars are as labeled.  Striations in the background show
how this plot would appear if the unmarginalized density had the form
$\rho(r)\propto 1+(r/r_0)^{-1.6}$.
}
\end{figure}

Figures~1 and~2 show examples of
what one can learn about galaxies' winds at high redshift from data that 
are currently available.  These data, described more thoroughly
in Adelberger et al. (2002), consist of measured redshifts
for 431 Lyman-break galaxies in 6 fields at $z\sim 3$  together
with high-resolution spectra of a bright background QSO in the
middle of each field.  
The leftmost panel of Figure~1
shows the mean Lyman-$\alpha$ transmissivity of the IGM as
a function of distance from Lyman-break galaxies at redshift $z\sim 3$.
Increases in the HI content of the IGM lead to more Lyman-$\alpha$
absorption and hence to a lower
Lyman-$\alpha$ transmissivity. (Here the IGM's
Lyman-$\alpha$ transmissivity at redshift $z$ is quantified
as the ratio of observed flux in a QSO's spectrum at wavelength
$\lambda=(1+z)\lambda_{Ly\alpha}$ to the QSO's expected flux
if there were no absorption from the IGM.)  The figure shows
that as one approaches a Lyman-break galaxy from afar, the
density of intergalactic HI at first begins to rise.  This is in accord
with the simple view that galaxies ought to be found where the
density of matter (including HI) is highest.  But at
small separations ($r\la 0.5h^{-1}$ comoving Mpc)
something else happens; intergalactic
HI largely disappears.  One interpretation is that the galaxies'
winds have largely driven away all material within this radius;
the competing hypothesis that the galaxies' light has ionized
the HI is ruled out by the large size of region with lowered
HI content (Adelberger et al. 2002).  

The right panel of Figure~1
shows the difference between the neutral hydrogen content of 
the intergalactic medium in young galaxy clusters and in
young voids.  Far more neutral hydrogen is found in the young
clusters.  The trend appears stronger than would be expected
solely from the fact that galaxy clusters ought to
contain more matter of all sorts (Adelberger et al. 2002).
A possible explanation is that the roiling of the young
intracluster medium by galaxies' blast-waves creates density
inhomogeneities and increases hydrogen's neutral fraction
(Adelberger 2002, 2003).

If Mpc-scale winds are responsible for the HI results just
discussed, one might expect to find an increase in 
the density of metals near galaxies.
This appears to be the case, as can be shown in a variety
of ways.  Consider first the triptych in the left half of Figure~2.
Shown are the two dimensional correlation functions of
galaxies with galaxies, of galaxies with CIV absorption systems,
and of galaxies with Ly-$\alpha$ transmissivity decrements.
The panels show how the densities of HI, CIV, and other galaxies
vary with spatial separation from a typical Lyman-break
galaxy at redshift $z\sim 3$; one should imagine that
the typical galaxy is at the origin and that the shadings
indicate the amount of material at various angular and
redshift separations.  $\xi({\mathbf r})$ is
defined so that the mean density at any separation ${\mathbf r}$ is
$\rho({\mathbf r}) = \langle\rho\rangle (1+\xi({\mathbf r}))$
with $\langle\rho\rangle$ the global mean.
Positive values of $\xi$ correspond to overdensities, negative
to underdensities.  One may immediately see that the intergalactic
medium near galaxies at $z\sim 3$ contains a large overdensity
of CIV systems.

Now consider the right panel of Figure~2,
which shows the same data marginalized in the redshift direction,
$\bar\rho(r_\theta, <r_z)\propto\int_0^{r_z}dr_z'\rho(r_\theta, r_z')$,
where the marginalization helps remove dependence on
redshift measurement errors or peculiar velocities.
Contemplation of this figure reveals
(a) that the ratio of CIV to HI density increases 
near galaxies,  suggesting perhaps that the intergalactic metallicity is highest 
close to galaxies, and (b) that on large scales the
galaxy-CIV cross-correlation function is similar to the galaxy-galaxy
auto-correlation function, suggesting that CIV absorption systems and galaxies
may be similar objects.
Readers will find a slightly less glib discussion in Adelberger et al. (2002).

\section{Ideal data}
I will not pretend that the data just presented are entirely
convincing.  In fact the entire point of my talk is to
advertise their inadequacy!  Some might worry that the statistical significance
of any one of these results is low, but that does not concern
me much.  The flaw is easily fixed; existing telescopes will shrink
our error bars by a factor of a few in a relatively short time.
I am more concerned with the fundamental limitations of the data.
There are several.  I will mention two.

Measuring the content of the IGM along a single skewer through
the galaxy distribution can tell us something about the
characteristic stalling radii of the winds, and hence about their
typical energies, but it tells us almost nothing about their geometry.
Are they (e.g.) isotropic or bipolar?  This has significant implications
but cannot be determined if the IGM is observed only along a single skewer.
Ideally one would have several background QSOs whose light pierced
each galaxy's wind at a range of angles and impact parameters.

More serious is the fact that working at $z\ga 3$ forces us to rely
almost exclusively on the IGM's CIV content as a proxy for its metal
content.  The density of CIV is not related in a simple way
to the metal content of the IGM (see, e.g., Adelberger et al. 2002).  Inferences
about intergalactic metals based solely on CIV will always be somewhat
suspect---to say the least.  But metals are the most direct signature of supernovae
,
and measuring the dispersal of metals out of galaxies and into the IGM 
will be central to our attempts at understanding supernova-driven
winds.   Ideally one would be able to measure the density of
several metallic species, not merely CIV.  This would provide constraints
on the ionization parameter and let one say something definitive about metallicity.
The relative abundances of different elements would also (in principle)
provide information about their nucleosynthetic origin.  But the vast
majority of the strongest absorption lines at intergalactic temperatures and densities
have wavelengths similar to or shorter than Lyman-$\alpha$'s.  This means that
at redshifts $z\ga 3$ they are buried in the thick Lyman-$\alpha$ forest
and are extraordinarily difficult to detect.  Only at redshifts $z\la 3$, where
the decreasing density of the universe slows recombination and thins the forest,
can these metallic absorption lines be straightforwardly found.

So this, then, is my picture of the ideal survey for quantifying empirically
how supernova winds have affected the universe.  Find a field with
a very large number of QSOs at redshifts $1\la z\la 3$ and obtain
high-S:N, high-resolution spectra of each.  (The upper redshift bound
comes from our desire that several intergalactic metal lines 
be easily detectable, the lower bound from the fact that
the comoving density of star-formation, and presumably the prevalence
of powerful winds, declines rapidly at $z<1$.)  Derive from the QSOs' absorption
lines a detailed picture of the distribution of HI and metals in
the intergalactic medium.  Conduct a redshift survey of galaxies at $1\la z\la 3$
in the same field and discover where star-forming galaxies lie amid the jumbled distribution of
intergalactic matter.   Then compare the relative spatial distributions
of galaxies and intergalactic material.  To the extent that
galaxies send winds into their surroundings, one should
find that the intergalactic medium near them is disturbed.  Perhaps
most of the intergalactic gas will have been driven away; perhaps
large amounts of metals will be found at the winds' stalling radii. 
In any case, one should be able to answer the
following questions (and others):
What fraction of the energy released by supernovae appears
to have been imparted to galaxies' winds?  What sorts
of galaxies drive winds?  Are winds from dwarf galaxies really the
most important, as many have argued?
Are the winds more isotropic or bipolar?  What is their dominant
effect on the the large fraction of the universe's baryons that
lies outside of galaxies?  And so on.  

\begin{figure}
\plottwo{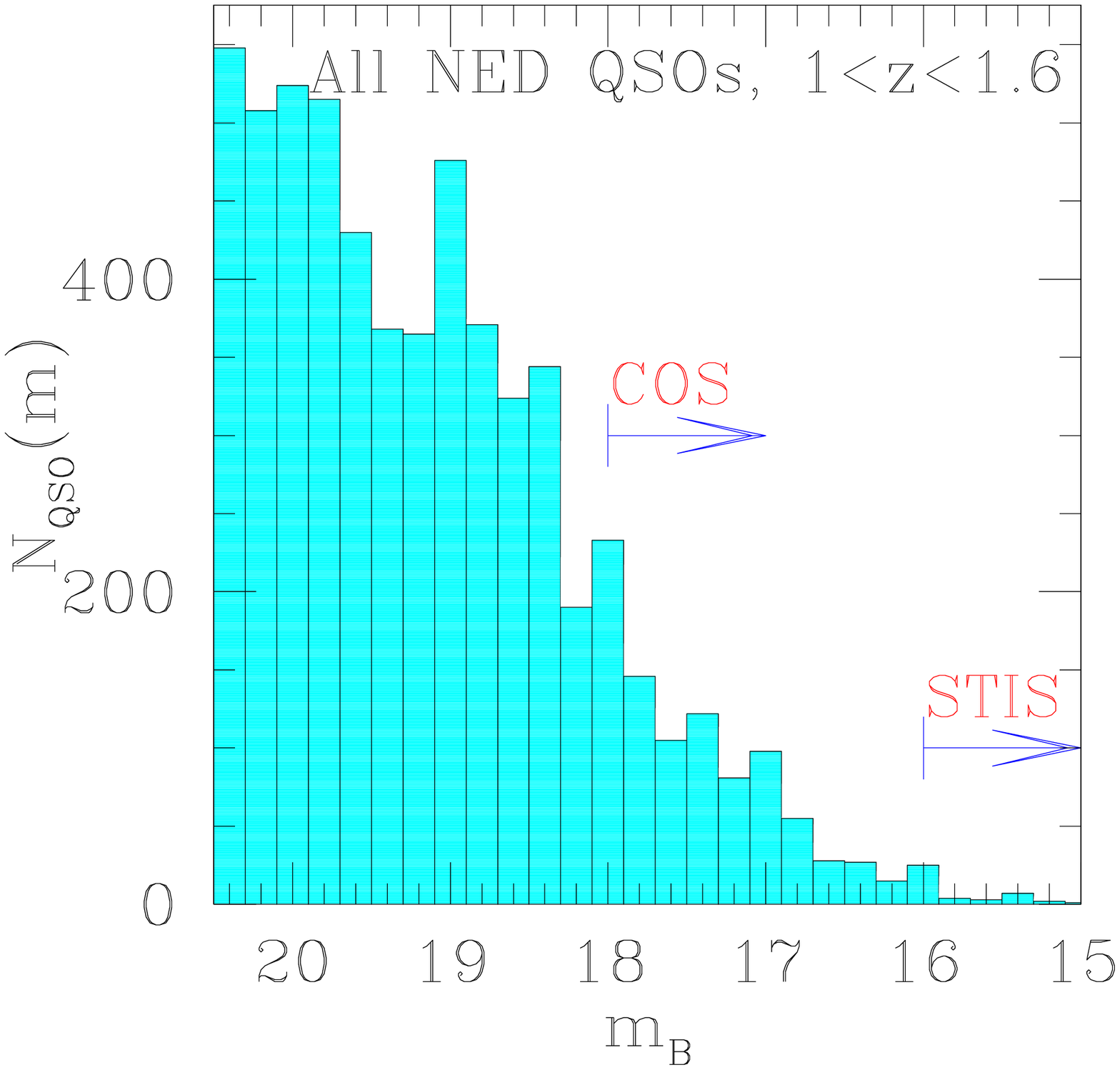}{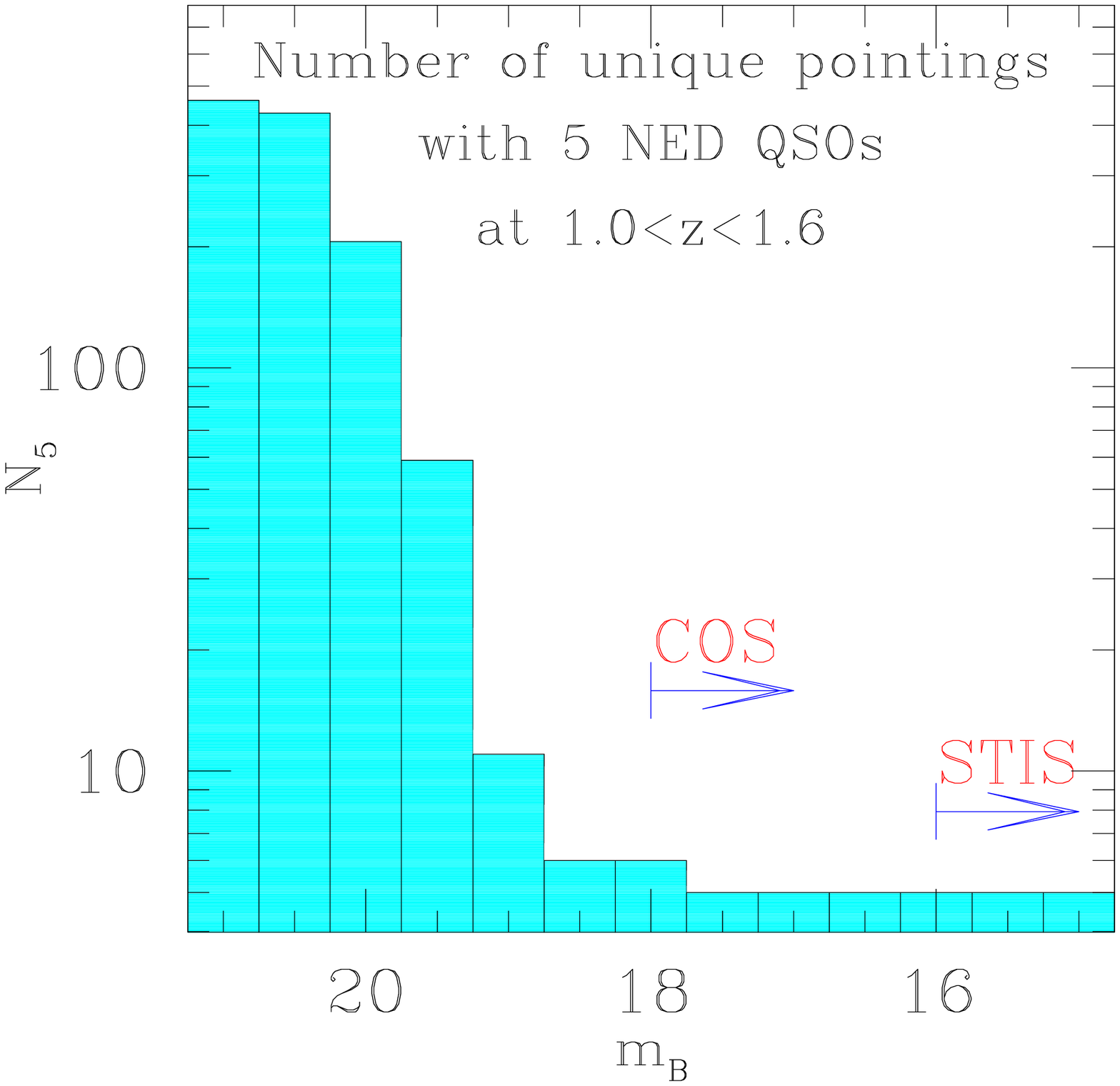}
\caption{Left:  the number of QSOs with $1.0<z<1.6$ in NED
(including the Sloan EDR) as a function of apparent magnitude.
Also shown are the rough magnitude limits for high S:N,
high resolution spectroscopy with STIS and COS.
NED is seriously incomplete at the faintest magnitudes, and
the true number density of QSOs rises much more rapidly at
faint luminosities than this plot suggests.
Right:  the number of unique $r=20'$ pointings that contain
5 or more of the NED QSOs in the left panel as a function
of the imposed QSO magnitude limit.  STIS and COS will
let us obtain sampling this dense of the HI and metals
in the IGM for only a handful of pointing across the entire
sky---and the ideal number of QSO in a field this size
is much higher than 5.  With a somewhat deeper magnitude
limit (e.g., $m\sim 20$) dense sampling of the IGM would
be vastly easier.  The gain is much larger even than this
figure suggests because of NED's incompleteness at fainter
magnitudes.  See text.
}
\end{figure}

Now let us think how such a survey might be constructed.  Obtaining galaxy
redshifts in this redshift range is trivial with 8m-class telescopes (e.g., 
Steidel et al. 1996, Adelberger 2002).
That part of the problem is solved.
The difficulty lies in obtaining quality spectra of the background QSOs.
Over much of the redshift range $1\la z\la 3$, Lyman-$\alpha$ and (especially) the strongest metal
lines lie in the observed near- or far-UV and can only be observed from space.
Unfortunately only a small number of QSOs are bright enough to be observed
with current and planned space-based UV spectrographs.  As an illustration,
the left panel of Figure~3 compares the rough high-resolution, high-S:N
magnitude limits of STIS and COS to the apparent magnitude distribution of
all known QSOs in the (arbitrary) redshift range $1.0<z<1.6$.  Only a small fraction
of these QSOs are bright enough to observe with either spectrograph.
The few hundred sources accessible to COS may seem like more than enough, but this
is not true, at least not if the goal (as it should be) is to obtain information
about the intergalactic distribution of hydrogen and metals along numerous
skewers through the foreground galaxy distribution.  This is illustrated
by the right panel of Figure~3, which shows,
as a function of assumed magnitude limit, 
the number of unique pointings in the entire sky that contain~5
or more NED QSOs at $1.0<z<1.6$ within a circle of radius $r\sim 20'$.
The number of QSOs and the circle radius were selected arbitrarily;
the plot is designed only to illustrate how much easier it is to find fields
with large numbers of background sources as the QSOs' limiting magnitude
becomes fainter.  Because NED is increasingly incomplete towards fainter
magnitudes, the true gains from a fainter magnitude limit are far larger even than
those indicated by this plot.

There are evidently a few known pointings with 5 or more QSOs suitable
for COS within a $r=20'$ circle.  Might these pointings be sufficient
for the project we are advocating?  The answer, sadly, is no.  5 QSOs
per $20'$ circle works out to roughly one QSO per $\sim 250$ square
arcminutes.  For an $\Omega_M=0.3$, $\Omega_\Lambda=0.7$ cosmology
at $z=1.5$, the distance along the plane of the sky between the QSOs in
these pointings will typically be more than $10h^{-1}$ comoving Mpc.
For comparison, the winds (apparently) detected around galaxies
at $z\sim 3$ have a radius smaller than $1h^{-1}$ Mpc.  If our goal
is to have a few QSOs behind an average galaxy's expanding blast-wave
(in order, e.g., to say something about the blast-wave's geometry, or learn
about the connection between galaxy properties and blast-wave properties
on a case-by-case basis),
we will need a density of background QSOs that is at least 100 times higher.

A crude estimate of the required spectrographic sensitivity
comes from the following calculation, which could
and should be improved.  Assume the QSO luminosity distribution
at the bright end has the shape $n(l)\propto l^{-\alpha}$ with $\alpha\simeq 3.7$
(e.g., Pei 1995).
Then the number density of QSOs brighter than magnitude limit
$m$ is proportional to $10^{0.4(\alpha-1)m}$, and an improvement
in sensitivity of 2 magnitudes over COS would lead to the desired factor of 100 increase
in the density of background QSOs.  



The good news is that a gain in sensitivity of this size
is hardly inconceivable.  The required increase 
in sensitivity over COS+HST is similar to the increase
that will come when STIS is replaced by COS, for example. 
Obtaining the data set that I have described, and settling
at last the vexing question of how supernovae have changed the
evolution of the baryonic universe, is not much beyond our reach.
How long will we have to wait?  This is the bad news.
The necessary QSO absorption lines are almost all at wavelengths
that can only be observed from space; the atmosphere is an obstacle
that will not change.  Under current NASA plans these
data will not be available until most of us are retired or dead.

\bigskip
\bigskip

I would like to acknowledge the major contributions of my collaborators
Chuck Steidel, Alice Shapley, and Max Pettini, the support and
patience of this meeting's organizers, and the hospitality of
Bukowski's, in Boston, where much of this polemic was composed.


\begin{references}
\reference Adelberger, K.L. 2002, PhD Thesis
\reference Adelberger, K.L., Steidel, C.C., Shapley, A.E., \& Pettini, M. 2002,
	\apj, submitted
\reference Adelberger, K.L. 2003, \apj, submitted
\reference Cen, R. \& Bryan, G. L. 2001, \apjl, 546, 81
\reference Kaiser, N. 1991, \apj, 383, 104
\reference Pei, Y.C. 1995, \apj, 438, 623
\reference Pen, U.-L. 1999, \apjl, 510, 1
\reference Ponman, T.J., Cannon, D.B., \& Navarro, J.F. 1999, Nature, 397, 135
\reference Springel, V. \& Hernquist, L. 2002, \mnras, in press
\reference Steidel, C.C.,  Giavalisco, M., Pettini, M., Dickinson, M., \&
        Adelberger, K. L. 1996, \apjl, 462, 17
\reference Weil, M. L., Eke, V. R., \& Efstathiou, G. 1998, \mnras, 300, 773
\reference White, S.D.M. \& Rees, M.J. 1978, \mnras, 183, 341
\end{references}
\end{document}